\documentclass[twocolumn,showpacs,nofootinbib,tightenlines,nobibnotes, aps, amssymb,amsmath,prb]{revtex4}
\usepackage{graphicx}% Include figure files
\usepackage{epsfig}% Include figure files

\newcommand{\la}{\langle}
\newcommand{\ra}{\rangle}

\newcommand{\be}{\begin{eqnarray}}
\newcommand{\ee}{\end{eqnarray}}

\begin{document}
%\preprint{APS/123-QED}
%draft
%\twocolumn[\hsize\textwidth\columnwidth\hsize\csname @twocolumnfalse\endcsname
\title{Spin nematics and magnetization plateau transition in anisotropic Kagome magnets}
\author{Kedar Damle$^1$ and T. Senthil$^{2,3}$}
\affiliation{%
$^1$ Department of Theoretical Physics,
Tata Institute of Fundamental Research,
1, Homi Bhabha Road, Colaba, Mumbai 400005, India\\
$^2$ Center for Condensed Matter Theory, Indian Institute of
Science, Bangalore 560012, India \\
$^3$ Massachusetts Institute of Technology, Cambridge, MA 02139,
USA}

\date{Feb 28, 2006}

\begin{abstract}
We study $S=1$ kagome antiferromagnets with
isotropic Heisenberg exchange $J$ and strong easy axis single-ion
anisotropy $D$. For $D \gg J$, the low-energy physics can
be described by an effective $S=1/2$ $XXZ$ model with antiferromagnetic
$J_z \sim J$ and ferromagnetic $J_\perp \sim J^2/D$. Exploiting this connection, we argue that
non-trivial ordering into a ``spin-nematic" occurs whenever $D$ dominates over
$J$, and discuss its experimental signatures. We also study a magnetic
field induced transition to a magnetization
plateau state at magnetization $1/3$ which breaks lattice translation
symmetry due to ordering of the $S^z$ and occupies a lobe in the $B/J_z$-$J_z/J_\perp$
phase diagram.

\end{abstract}

\vskip2pc

\maketitle

\begin{figure}[t]
\includegraphics[width=\hsize]{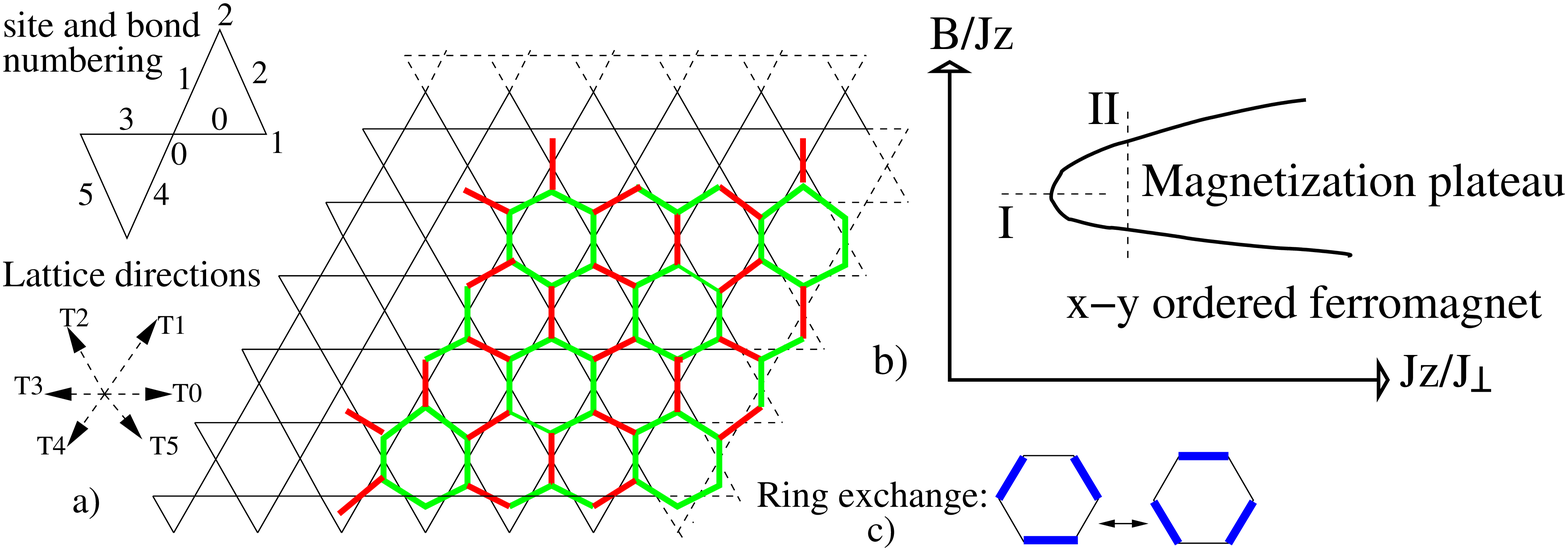}
\caption{(color online) (a) Periodic Kagome lattice and the honeycomb net whose
bonds pass through the kagome sites. In the plaquette ordered state,
red honeycomb edges have no dimer ($\sigma^z=+1$), while green hexagons resonate via
the ring-exchange process (shown in (c)). In the alternate columnar state at the same wavevector, dimers
cover all red edges ($\sigma^z=-1$) but not green ones. (b) Schematic phase diagram, showing
the scans I and II discussed in text.}
\label{lattice}
\end{figure}

{\it{Introduction:}}
Magnets with {\em geometrical frustration} (competition between different
spin exchange interactions caused by the lattice geometry) exhibit many interesting properties including spin-liquid like low temperature phases and unusual
spin correlations\cite{Moessnerreview,exptreview1}.
Kagome lattice (see Fig~\ref{lattice}) magnets provide many examples of this,
and several realizations with spins ranging
from $S=1/2$ to $S=5/2$\cite{otherkagome0,otherkagome1,otherkagome2} have been experimentally studied.
On the theoretical side, numerical
and analytical work suggests that
$S=1/2$ isotropic Heisenberg antiferromagnet on the kagome lattice is in an unusual
phase with an anomalously large density of
singlet excitations\cite{exactdia1,nikolic1} at $B=0$. At finite $B$, there also exists evidence for the presence of a robust magnetization plateau
state with magnetization pinned to $1/3$ of the saturation moment\cite{exactdia2}.

A particularly interesting example of frustrated magnetism is
provided by the Ni$^{2+}$ based effective $S=1$ Kagome magnet Ni$_3$V$_2$O$_8$, in which
isotropic exchange interactions compete along
with sizeable single-ion anisotropy terms (and weak Dzyloshinski-Moriya interactions) resulting
in a rich phase diagram in the presence of a magnetic field\cite{nickel1,nickel2}.
Motivated in part by this, we consider a kagome lattice model with
nearest neighbour antiferromagnetic spin exchange interaction
($J>0$) between spin $S=1$ ions in the presence of an easy-axis
single-ion anisotropy ($D>0$) along the $z$ axis with the
Hamiltonian
\be H &=& J\sum_ {\la ij\ra} \vec{S}_i \cdot \vec{S}_j
- D\sum_i (S^z_i)^2 - B \sum_i S^z_i \; , \label{H_S1} \ee
where
$\la ij\ra$ refer to nearest neighbour links of the
two-dimensional kagome lattice.

Unlike the {\em unfrustrated case}, the $D$ term may have important effects in frustrated
systems even if not very big.  We therefore study the limit of large $D/J$ and
show that interesting physics emerges: We
show that the ground state at $B=0$ is a quantum spin nematic associated
with ordering of $<(S^+)^2>$ without ordering of the spin itself.
Upon increasing the field, magnetization plateaus appear at
specific magnetization values. Of particular interest is a plateau
at magnetization $1/3$ which we show breaks translational
symmetry. The corresponding plateau transition has a number of
interesting properties which we discuss.

When $D/J$ is large and positive and $B \lesssim J$, each spin is predominantly in
the $m_z = \pm 1$ states, and we can describe the
low energy physics in terms of an effective Hamiltonian for
(pseudo-) spin $S=1/2$ variables $\sigma^z$. Explicit calculation to second order in $D/J$ yields the
following effective low energy Hamiltonian in this
regime\cite{footer1}: \be H_{\mathrm{eff}} &=& -\frac{J_\perp}{4}
\sum_ {\la ij\ra} (\sigma^+_i \sigma^-_j + h.c.) + \frac{J_z}{4}
\sum_ {\la ij\ra}\sigma^z_i \sigma^z_j - B \sum_i \sigma^z \; ,
\nonumber \label{H_S1/2} \ee Here, the $\vec{\sigma}$ are the
usual pauli spin matrices, and the parameters of $H_{\mathrm{eff}}$
are given by  $J_z \approx 4J + J^2/D$ and $J_\perp \approx
J^2/D$; thus, for large $D/J$ we
have $J_z/J_\perp \approx 4D/J + 1 + O(J/D)$. Clearly, the ground-state of this pseudospin $S=1/2$ $XXZ$ model for small
$J_z/J_\perp$ (which is not directly related to the physics of our
original $S=1$ problem) must be a ferromagnet polarized in the
$xy$ plane. Below we analyze the large $J_z/J_\perp$ regime
(appropriate for the large $D$ physics of the original $S = 1$ model)
separately for $B=0$ or small, and $B \sim J$.

When $B=0$, the dominant diagonal interaction $J_z$ leads to
frustration since it is impossible to have all pairs of
neighboring spins pointing anti-parallel to each other along the
$z$ axis on the kagome lattice. The ground state then lives
entirely in the highly degenerate minimally frustrated subspace
with precisely one frustrated bond (parallel spins)
per triangle, and is selected by the spin-exchange dynamics ($J_\perp$). This physics in the present $J_\perp
> 0$ case can be understood straightforwardly by thinking in terms
of variational wavefunctions (as was done
recently\cite{dariush,arun} on the triangular
lattice): Since the
spin-exchange $J_\perp > 0$ is {\em unfrustrated}, a good
variational wavefunction for the small $J_z/J_\perp$ ferromagnet
is simply $| \Psi_F \rangle = \Pi_i |\sigma^x_i = +1 \rangle$. Furthermore,
a natural description for the state at {\em
large} $J_z/J_\perp$ can be obtained by projecting $|\Psi_F
\rangle$ to the minimally frustrated subspace described above.
Since this subspace admits considerable
fluctuations in the values of $\sigma{_z}$, such a projected wavefunction
$| \Psi_\infty \rangle $ continues to gain `kinetic
energy' from spin-exchange processes, while minimizing
the diagonal interaction energy by construction.

Thus, $x$-$y$ ferromagnetic order {\em persists even in the large
$J_z/J_\perp$ limit} at $B=0$, and this remains valid for small
$B$ as well. Moreover, $\sigma^z$ correlators in $|\Psi_\infty
\rangle$ are simply given by the $T=0$ correlations of the
classical Ising model on the Kagome lattice, and their
short-ranged nature\cite{clkag} rules out any co-existing
$\sigma^z$ spin density wave order. [The same conclusion has been
reached recently in other ways\cite{sengupta0} and confirmed
numerically\cite{kdun1}.] What does this analysis imply for the
original $S = 1$ magnet? As the pseudospin operator $\sigma^+ \sim
(S^+)^2$, the $xy$ ferromagnet of the pseudospin magnet actually
corresponds to an $xy$ {\em spin nematic} state where $<(S^+)^2>
\neq 0$ but $<\vec S> = 0$. Thus, we conclude that spin-$1$ Kagome
magnets with strong easy axis anisotropy order into such a spin
nematic phase with $<(S^+)^2> \neq 0$ for $B=0$ and its immediate
vicinity.

The presence of this nematic ordering is one of our main
conclusions.  As a state that breaks the global $U(1)$ symmetry of
spin rotations about the easy axis this nematic will have a
gapless linear dispersing `spin' wave which will lead to a $T^2$
contribution to the low temperature specific heat. Further this
state will have a non-zero finite spin susceptibility for fields
both parallel and perpendicular to the easy axis. Despite these
similarities with conventional ordered antiferromagnets there will
not be any magnetic Bragg spots in neutron scattering as the spin
itself is disordered.

In passing we note that the same considerations on a triangular
lattice again predict nematic ordering which {\em coexists} with
spin density wave ordering of the $z$-component of the spin - this
follows directly from the arguments above and the results of Ref.
\onlinecite{arun,dariush,wessel}.

Returning to the Kagome lattice, as we turn on a magnetic field
$B$, the magnetization will initially rise smoothly with field
since the nematic persists for small $B$. As the field is
increased
 to $B \sim J$, there will be
plateaus where the magnetization is field independent and fixed to
specific commensurate values of the magnetization. We now show
that for a range of $B$ away from $B=0$ there is such a plateau
state at magnetization $1/3$ where the ground state is a
lattice-symmetry broken spin-density wave (SDW) state (in which
the $z$ component of the spins order as in Fig~\ref{lattice}).

Working again with the effective
 $XXZ$ pseudospin Hamiltonian we
begin in the extreme limit of $J_\perp/J_z \rightarrow 0$ by writing $B$ in terms of a
reduced field $b$ as $B = J_z b$ and noting that the $z$ coupling
and field terms in $H_{\mathrm{eff}}$ can be combined and rewritten
as $\frac{J_z}{8}\sum_{t} (\sigma^z_t - 2b)^2$, where the sum is now over all triangles $t$ of the kagome lattice.
The physics in this (classical) limit is now clear: For $0<b<1$, the energy is minimized by having
two of the spins in each triangle pointing up and one pointing down, which yields a
magnetization
equal to $1/3$ of the saturation magnetization, while for $b>1$, the ground state magnetization
is locked to the saturation value by having all spins pointing up. Thus, one expects a magnetization plateau at
$1/3$ of the saturation magnetization in the vicinity of $b=0.5$, where the energy gap
to change in magnetization is largest. In this (classical)
limit, the ground state has extensive degeneracy, as may be easily seen by noting that
the manifold of low-energy configurations can be mapped to the perfect dimer covers
of the honeycomb lattice whose edges pass through the kagome sites (with
each down spin corresponding to a dimer covering the corresponding honeycomb edge).

Let us now turn on a small $J_\perp$. Apart from an unimportant
constant shift in energy, the leading non-trivial effect of this perturbation is easily
seen to arise at third order in degenerate perturbation theory and correspond
to a `ring-exchange' term which allows flippable hexagonal plaquettes to resonate with amplitude $t \sim -J_\perp^3 /J_z$ (Fig~\ref{lattice} (c)). This quantum dimer model on the honeycomb
lattice is known to be in a crystalline `plaquette' state that breaks the lattice translation
symmetries of the honeycomb lattice in order to maximize the number of independently flippable
plaquettes from which the system can gain kinetic energy\cite{moessnerhoney}. This implies a ground state with long-range density wave order of
the $\sigma^z$ and of the bond energies $\sigma^+_i \sigma^-_j + h.c.$ (Fig~\ref{lattice}). \begin{figure}[t]
\includegraphics[width=\hsize]{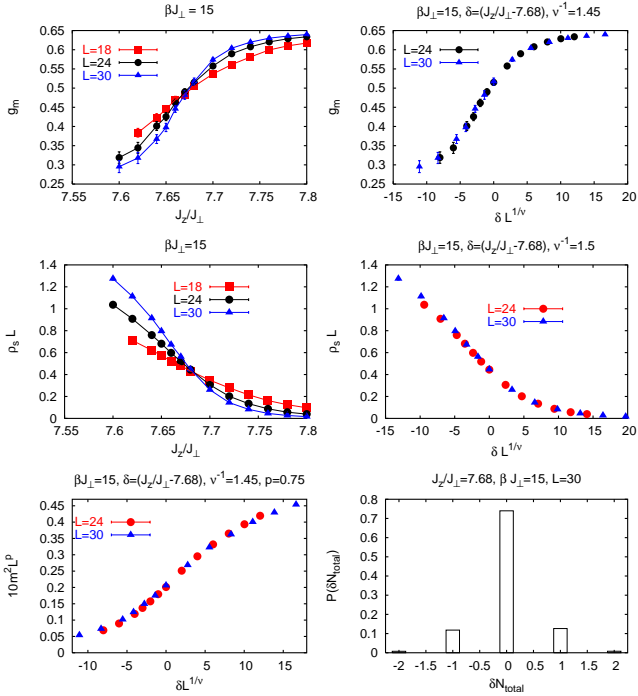}
\caption{ Numerical evidence for direct second order transition at the particle-hole symmetric tip of
magnetization plateau lobe (scan I). Here $m^2=S_{\rho}^{00}(Q,\omega_n=0)/4L^2\beta$,
while $g_m = 1-\langle m^4 \rangle/(3\langle m^2 \rangle^2)$ is the standard Binder
cumulant of the spin-density wave order parameter and $\delta N_{tot} \equiv 0.5 \delta \sigma^z_{tot}$}
\label{rhosandnq}
\end{figure}

Thus, the plateau state is stable for large finite $J_z/J_\perp$ and is therefore expected to
occupy a lobe in the $B/J_z$-$J_z/J_\perp$ plane
(Fig~\ref{lattice}).
Clearly, the tip of this lobe represents a special point along the
locus of plateau transitions as the vicinity of the tip is
distinguished by the presence of low energy `particle-hole'
symmetry corresponding to equal energies for `quasiparticle' and
`quasihole' excitations (here quasiparticles and quasiholes are
distinguished by the sign of the magnetization deviation from
$1/3$ that they induce by their presence). Given that the plateau
state breaks lattice translation symmetry, the transition to the
ferromagnet (nematic) at the tip and away presents interesting
possibilities: Conventional Landau theory would predict either an
intermediate phase with both orders present or a first order
transition. However recent work\cite{Senthil0} has shown that
Landau theory itself can fail in closely related bosonic models -
the result in such cases is expected to be an unusual direct
second order phase transition.

{\it Numerical results:} The foregoing provides the motivation for
our numerical study of the $S=1/2$ XXZ model at large $J_z/J_\perp$ and finite field $B \lesssim 0.5 J_z$.
We  use the well-documented stochastic series expansion (SSE)
QMC method\cite{Syljuasen0} to access the phase diagram.
(At large values of $J_z/J_\perp$, some modifications developed recently\cite{gros,dariush}
were used to improve the algorithmic efficiency).
Most of our data is on $L \times L$ samples (where $L$ is number of unit
cells) with periodic
boundary conditions and $L$ a multiple of six ranging from 18 to 30
at inverse temperatures $\beta$ ranging from $5 /J_{\perp}$ to $15/J_\perp$.
We use standard SSE estimators\cite{Syljuasen0} to calculate the ferromagnetic
stiffness $\rho_s$, the equal time ($C_{\rho}^{\alpha \alpha^{'}}(q,\tau=0) = \langle \sigma^z_\alpha(q) \sigma^z_{\alpha^{'}}(-q)\rangle$) and static correlators ($S_{\rho}^{\alpha \alpha^{'}}(\vec{q},\omega_n=0) = \int_{0}^{\beta}d \tau C_{\rho}^{\alpha \alpha^{'}}(\vec{q},\tau)$) of $\sigma^z_\alpha$,
and the static correlator of the `kinetic energy' $K_l = (\sigma^+\sigma^-
+h.c.)_l$ on link $l$  ($S_{\mathrm{K}}^{\alpha \alpha^{'}}(\vec{q},\omega_n=0) = \int_{0}^{\beta}
d \tau C_{\mathrm{K}}^{\alpha \alpha^{'}}(\vec{q},\tau)$) (here $\alpha$ and $\alpha^{'}$ refer
to the 3 basis sites and six bond orientations in a unit cell, and all site
and bond types shown in Fig~\ref{lattice} are assigned the coordinates of site type $0$ when defining the fourier transform).\begin{figure}[t]
\includegraphics[width=\hsize]{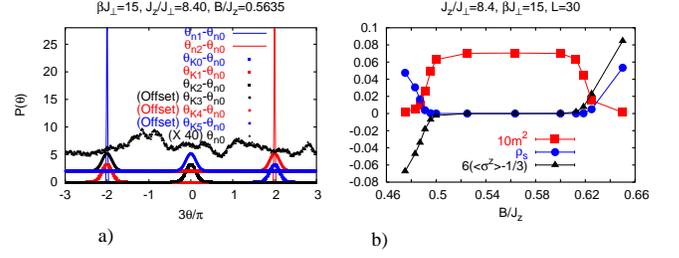}
\caption{a) Histograms of relative and absolute phase of all order parameters.
b) Vertical scan (II) showing plateau state at $J_z/J_\perp = 8.4$.}
\label{thetahistogram}
\end{figure}
%\begin{figure}[t]
%\includegraphics[width=.5\hsize]{vertscan.eps}
%\epsfxsize=.8\hsize \centerline{\epsfbox{binder.eps}}
%\caption{Scan (II) at constant $J_z/J_\perp = 8.4$ and $\beta J_\perp = 15$ for $L=30$ sample.}
%\label{vertscan}
%\end{figure}

By analyzing the $L$ and $\beta$ dependence of
the Bragg peaks at $\pm Q = \pm (-2\pi/3, 2\pi/3)$ (components refer to projections along
T$_0$ and T$_1$ (Fig~\ref{lattice})) seen in
the static correlation functions of $\sigma^z$ and $K_l$, we conclude that spatial order is established
at these wavevectors  when ferromagnetism is destroyed in the plateau state; the observed wavevector $Q$ is
the ordering wavevector of the plaquette and columnar states of Fig~\ref{lattice}. The static structure
factors near the onset of the plateau state also reveal the presence of an interesting `dipolar'
structure somewhat analogous to the dipolar part of dimer correlators in the classical honeycomb
lattice dimer model\cite{youngblood}. These seem to simply reflect the {\em local} magnetization
1/3 constraint imposed by the $B$ and $J_z$ terms in this region of parameter space\cite{senthilkedarunpub} and persist
across the transition into the ordered state.

To further probe the
nature of the ordering, we also measure the statistics of the phases $\theta_{n\alpha}$,
$\theta_{K\alpha}$ of nine complex order parameters $\psi_{n \alpha}=\sigma^z_\alpha(Q,\omega_n=0)$ and $\psi_{K
\alpha} = K_{\alpha}(Q,\omega_n=0)$. From Fig~\ref{thetahistogram} (a), we see that
that all relative phases are essentially pinned by the energetics of the plateau state and there is
one independent phase degree of freedom which we take to be $\theta_{n0}$ ($\theta_{n0} = 0,
\pm 2\pi /3$ correspond to the three equivalent plaquette ordered states
while $\theta_{n0} = \pi, \pm \pi/3$ correspond to the alternative
`columnar' states at the same wavevector). Clearly, this overall phase is very
weakly
pinned (compared to the relative phases) even at low temperatures relatively far from the critical region, but the presence of distinct
Bragg peaks in the kinetic energy correlator strongly suggest that the ordering is of
the plaquette type as simple caricatures of the columnar state contain no dimer resonances.

The statistics of these phases can be interpreted
in terms of a Landau free energy $F$ written in terms of the order parameters
$\psi_{n\alpha}$, $\psi_{K\alpha}$: $F = F_1(|\psi_{n\alpha}|^2,|\psi_{K\alpha}|^2) + F_2$, where $F_1 = r_1|\psi_{n\alpha}|^2
+r_2|\psi_{K\alpha}|^2 + u_1(|\psi_{n\alpha}|^2)^2 +u_2(|\psi_{K\alpha}|^2)^2 +\dots $ is
insensitive to lattice geometry and phase information and $F_2$ encodes phase information. When $F_1$ develops a minimum at non-zero values of its argument, $F_2$ determines
the details of the ordering pattern. The most important terms in $F_2$ are quadratic invariants which encode the
{\em local} magnetization $1/3$ constraint on each triangle (imposed by $J_z$ and $B \sim 0.5 J_z$), and the kinetic energy gain from having
pairs of neighbouring sites exchanging spins ($-J_\perp$). The former may be written as
$a(|\Psi_n \cdot I|^2+|\Psi_n \cdot \Omega^*|^2)-b|\Psi_n \cdot \Omega|^2$ with $a$ and $b$ both
expected to be positive on energetic grounds (we use the dot product
notation $v \cdot w \equiv \sum_\alpha v_\alpha w_\alpha$,
and $\Psi_n \equiv (\psi_{n0}, \psi_{n1}, \psi_{n2})$, $\Omega \equiv (1, e^{2\pi i/3}, e^{4\pi i/3})$, $I=(1,1,1)$),
while the latter is given by $c |\Psi_K \cdot \Theta|^2$ with $c$ expected to be negative on energetic grounds (here $\Psi_K \equiv (\psi_{K0},\psi_{K1},\psi_{K2},\psi_{K3},\psi_{K4},\psi_{K5})$ and
$\Theta \equiv (1, e^{4\pi i/3}, e^{2\pi i/3},e^{4\pi i/3},1,e^{2\pi i/3})$). In addition,
symmetries also permit a quadratic cross-term $\left [(\Psi_K \cdot \Theta)(\Psi_n^* \cdot \Omega^*)e^{-2\pi i/3} + h.c.\right ] $. Landau theory thus predicts that the relative phases are such that the natural order parameters
$\Psi_K \cdot \Theta$ and $\Psi_n \cdot \Omega$ have maximum modulus and relative phase
of $2\pi/3$ (assuming the coefficient of the cross-term is negative), and clearly
this correctly reproduces
the phase relationships exhibited by the data (Fig~\ref{thetahistogram} (a)).
Finally, the absolute phase
is expected to be chosen by cubic terms of the form $Re\left [ (\Psi_n \cdot \Omega)^3 \right ]$, and
our data suggests that the {\em renormalized} value of the corresponding coefficient is extremely small
even away from the phase boundary.

We have also studied the nature of the phase boundary between the plateau state and
the ferromagnet by performing several scans, of which we show data for two
here. The first scan (scan I shown in Fig~\ref{lattice}) at constant $B/J_z=0.5635$ was chosen as it intersects the
phase boundary at a particle hole symmetric point which we identify as the tip of the plateau lobe. The
sharpness of the crossings seen in the plots (Fig~\ref{rhosandnq}) of the Binder cumulant
$g_m$ and of $\rho_s L$ for different sizes strongly
suggest that we have reached the asymptotic low temperature regime and provide indications that
the transition is a direct second order transition at $J_z/J_\perp \approx 7.68 \pm 0.02$ with
$z = 1$. From a scaling collapse of these crossing curves, we estimate
$1/\nu \approx 1.45 \pm 0.2$, while similar analysis for the order parameter $m^2$
gives $ 2 \beta/\nu \approx 0.75 \pm 0.1$; however, we emphasize that
these are {\em estimates} and a more detailed study with much larger sizes is needed to definitively
rule out a very weak first order jump or small coexistence region and obtain precision values of exponents.
The second, vertical scan (scan II) was performed at $J_z/J_\perp = 8.4$ primarily to confirm the
existence of the plateau
state over an appreciable range of $B$, and yields a plateau state for $0.49 \lesssim B/J_z \lesssim 0.62$
(Fig~\ref{thetahistogram} (b)).
Near the transition points, no clear evidence of a first order jump is seen, nor is there a
well-resolved region with coexisting order parameters (further details regarding the nature of these transitions
will be discussed separately\cite{senthilkedarunpub}).

{{\it Summary}}: To summarize, we have demonstrated that $S=1$ kagome antiferromagnets with moderately
strong single-ion anisotropy of the easy-axis type exhibit an interesting spin-nematic state
at and in the vicinity of $B=0$, as well as a lattice-symmetry broken spin-density wave
magnetization plateau state at $1/3$ magnetization for $B \sim 0.5 J$.
We have also presented numerical evidence that
the transition between these is of an unusual direct second order type at least at the tip
of the plateau lobe in the $B/J$-$D/J$ plane. We hope our work will provide some impetus to look for these
effects in anisotropic $S=1$ kagome magnets.

{\it{Acknowledgements:}} We would like to acknowledge useful
discussions with A. Vishwanath D.~Dhar and A.~Paramekanti,
and assistance of A.~Sen and
S.~Mukherjee in preparation of figures. Computational resources of
TIFR as well as those kindly made available by V. Shenoy at IISC
Bangalore are gratefully acknowledged. TS acknowledges funding
from the NEC Corporation, the Alfred P. Sloan Foundation and an
award from the Research Corporation. During completion of this
work we became aware of parallel work\cite{Melkounpub} with
partial overlap, and would like to acknowledge useful discussions
with one of the authors (Y.B. Kim).

\end{document}